%% file: cryFundamentalFrequency.tex
\documentclass[journal]{IEEEtran}
\usepackage{amssymb,amsmath}
\usepackage{graphicx}
\usepackage{cite}
\usepackage{subfigure} 
\input {myshorts}

\title{Estimation of Infants' Cry Fundamental Frequency using a Modified SIFT algorithm}
\author{Dror~Lederman  \\
        Department of Radiology \\
        University of Pittsburgh, \\
        3362 Fifth Avenue, Pittsburgh, PA 15213  \\
        Tel. +412-641-2581, Fax +412-641-2582 \\
        E-mail:ledermand@upmc.edu}

\begin{document}

\maketitle

\newpage

\IEEEpeerreviewmaketitle

\begin{abstract}
     This paper addresses the problem of infants' cry fundamental frequency 
     estimation. The fundamental frequency is estimated using 
     a modified simple inverse filtering tracking (SIFT) algorithm. The performance of the modified SIFT is studied using a real database of infants' cry. It is shown that the algorithm is capable of overcoming the problem of under-estimation and over-estimation of the cry fundamental frequency, with an estimation accuracy of 6.15\% and 3.75\%, for hyperphonated and phonated cry segments, respectively. Some typical examples of the fundamental frequency contour in typical cases of pathological and healthy cry signals are presented and discussed. 
\end{abstract}

\IEEEpeerreviewmaketitle

\section{Introduction}
\label{sec:intro}
The fundamental frequency, often referred to as the pitch (which is actually the subjective perceptual estimation of the fundamental frequency by one's brain), is one of the most important features used to discriminate between various types of infants' cries \cite{Boukydis1985, Colton1981}. The fundamental frequency has been found to be a discriminative feature in severe medical problems such as SIDS \cite{Colton1981}, hypo/hyperglycemia \cite{Koivisto1974}, brain damage \cite{Boukydis1985}, cleft palate \cite{Michelsson1975}, hydrocephalus \cite{Michelsson1984}, cri-du-chat syndrome \cite{Vuorenkoski1966} and many others \cite{Boukydis1985}. Moreover, the fundamental frequency has been found to be directly correlated with diseases related to brain function at the early stages of child development \cite{Boukydis1985}. Consequently, accurate measurement of the fundamental frequency and the variation of the fundamental frequency along time is most important in order to obtain reliable deduction of information on the state of health of newborn babies. 

Several studies have addressed the  problem of cry fundamental frequency estimation (e.g. \cite{Fort1998, Mende2001, Escobedo2001, Wermke2002, Manfredi2006}). Fort and Manfredi \cite{Fort1998} proposed a method termed adaptive cepstrum which is based mainly on cepstral analysis with an optimal choice of the lifter length on each time-window in order to increase the frequency resolution. Escobedo et. al. \cite{Escobedo2001} used linear prediction (LP) analysis to estimate the fundamental frequency. Fort and Manfredi \cite{Fort1998}, as well as Mende \cite{Mende2001}, discussed some problems and pitfalls encountered when estimating fundamental frequency of cry signals. Parametric and non-parametric methods used to estimate the fundamental frequency and formants of newborn were analyzed and compared. These studies revealed the problem of identifying sub-harmonics of the fundamental frequency, in particular in hyperphonated cry segments.

\begin{figure}[htb]
  \centering
  \includegraphics[scale=0.75]{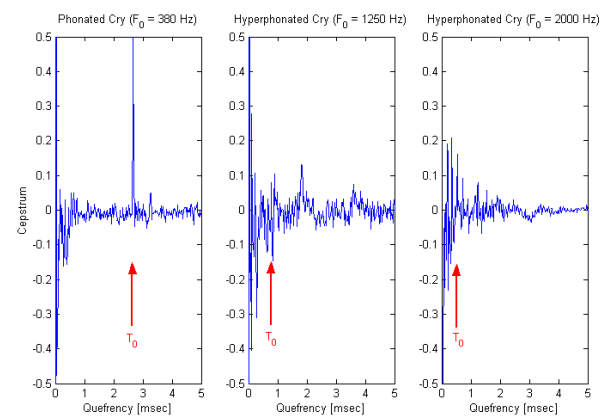}
	\caption{Cepstrum series for some typical cry segments. Left- phonated cry 
with $F_0 = $380Hz; Middle- hyperphonated cry with $F_0 = $1250Hz; Right- hyperphonated cry with $F_0 = $2000Hz  
(the arrows represent the 'true' fundamental frequency location estimated visually based on the time domain).}
	\label{fig:fig4}
\end{figure}

The vocal tract of newborn babies is associated with higher resonances and fundamental frequency than those of the adult. Infants' cry is mainly voiced and can be classified into phonated cry (a fundamental frequency of 250-700Hz) and hyperphonated cry (a fundamental frequency of 700Hz and above). In case of phonated cry segments, estimating the fundamental frequency is not significantly different than in the case of adults' speech. However, in case of hyperphonated cry segments, due to widening of the harmonic structure, separation of the excitation spectral contribution from that of the vocal tract might be very difficult. As a consequence, the use of cepstrum coefficients might fail to de-convolve the excitation source and the vocal tract. In order to illustrate this problem, fig. \ref{fig:fig4}  shows the cepstrum coefficients for some typical phonated and hyperphonated cry segments. The visually-estimated fundamental frequencies of the segments are around 380Hz, 1250Hz and 2000Hz, respectively. 

%\fig{fig4}{9.0cm}{8.0cm}{Cepstrum series for some typical cry segments. Left- phonated cry 
%with $F_0 = $380Hz, Middle- hyperphonated cry with $F_0 = $1250Hz, Right- hyperphonated cry with %$F_0 = $2000Hz  
%(The arrows represent the accurate fundamental frequency location).}
%      {fig:fig4}

In the phonated case, one can easily separate between the spectral contribution of the vocal tract and that of the excitation source. Hence, using a high pass lifter, one can easily estimate the fundamental frequency from the cepstrum as approximately 2.6msec, which corresponds to the visually-estimated frequency which is 380Hz. In both of the hyperphonated cases presented in fig. \ref{fig:fig4}, widening of the harmonic structure is clearly seen. In these cases, it is very difficult to isolate the spectral contribution of the vocal tract and to estimate the fundamental frequency based on the cepstrum. Therefore, the cepstrum coefficients might not be the best choice to form the features space for cry representation, particularly when classification of different cry patterns is of concern.

Another problem which often occurs in cry fundamental frequency estimation is incorrect identification of sub-harmonics of the fundamental frequency as duplication of the dominant fundamental frequency, and consequently, incorrectly classifying the segment as an hyperphonated segment. Vice versa, under-estimation of hyperphonated fundamental frequency can occur, which results in an incorrect classification of the cry segment as a phonated segment. 

In this paper, a modified simple inverse filtering tracking (SIFT) algorithm \cite{Markel1972} for infants' cry fundamental frequency estimation is proposed and investigated. The paper proceeds as follows. Section \ref{sec:SIFT} presents the modified SIFT algorithm. Some typical results and performance evaluation appear in Section \ref{sec:RESULTS}. A summary and conclusions appear in Section \ref{sec:SUMMARY}. 

\section{The modified SIFT algorithm}
\label{sec:SIFT}
\begin{figure*}[htb]
  \centering
  \includegraphics[scale=1.0]{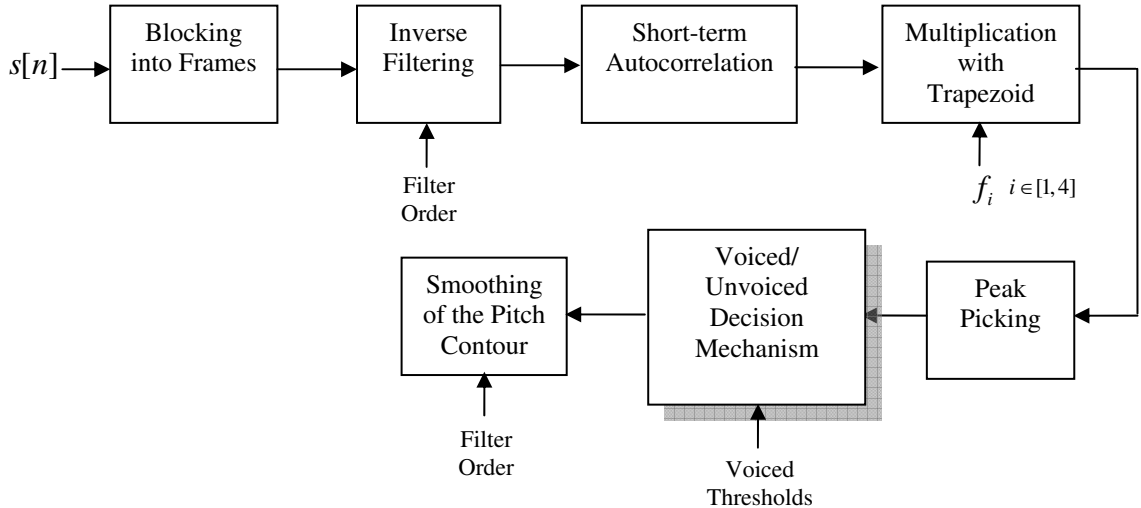}
	\caption{Block diagram of the modified SIFT.}
	\label{fig:fig3}
\end{figure*}

Fig. \ref{fig:fig3} presents a block diagram of the SIFT-based algorithm. The algorithm is performed as follows. In the first step, blocking of the signal into overlapping frames is performed. 
Since cry signals are non-stationary, short-term analysis must be used in which the signal is divided into quasi-stationary overlapping frames, and each frame is multiplied by a suitable Hamming window. The frame length is pre-chosen in order to ensure quasi-stationary of the framed signal on the one hand, and on the other hand, to ensure that each frame will include at least one period of the fundamental frequency, and to decrease the amount of computations as much as possible. The cry  fundamental frequency is known to have a wide range of possible values between 200Hz and 2500Hz. Consequently, the appropriate practical range for frames duration should be around 5-25 msec.

%\fig{fig3}{9.0cm}{8.0cm}{Block diagram of the modified SIFT.}
%      {fig:fig3}
%\figure[$1^{st}$ model]{\label{fig:hmm_a}\includegraphics[scale=1.0]{hmm_a}}

Next, glottal inverse filtering is performed. The purpose of inverse filtering is to attenuate the influence of the vocal-tract and to estimate the time excitation waveform. 
In order to ensure complete pre-whitening of the inverse filtering procedure, a proper choice of filter order according to the maximum number of formants possible in the given frequency range is necessary. Since typical values for the first three formants of cry signals are around 1kHz,3kHz and 4.5kHz \cite{Fort1998}, a reasonable choice for the filter order would be $M=4$, according to which either zero, one or two resonances can be accurately represented. 

Then, estimation of the autocorrelation sequence of the signal is performed. Considering a sampling frequency of 16kHz, in order to achieve good time resolution, 256 points are required (16 milliseconds) which correspond to more than 3 pitch periods. 

The next step is multiplying the autocorrelation by a trapezoid window in the frequency range defined by $f_i, i \in [1,4]$, in order to restrict the search to a reasonable cry fundamental frequencies range. 
Some typical examples are presented in fig. \ref{fig:fig1}. 
The fundamental frequency is then estimated using:
\begin{equation}
T_0 = \frac{1}{F_0} = \text{argmax}_{\eta} \left\{r_i(\eta) \right\},
\end{equation} 			
where $r_i(\eta)$   is the autocorrelation sequence multiplied by the trapezoid.
With the result, peak picking is performed in order to search for a candidate $F_0$, and a decision algorithm is applied in which the peak value is compared to a voiced threshold of 0.4. A peak which exceeds the threshold value of 0.4, or a borderline peak (between 0.3 and 0.4) with ``voiced history'' (last two frames are voiced) is defined as voiced. A peak which does not exceed the threshold value but has ``voiced neighbours'' (its preceding and its following segments are voiced) is also defined as voiced.  

The last step is smoothing of the resulting fundamental frequency contour. 
This is performed using a $4^{th}$-order median filter in order to depress abrupt changes in the fundamental frequency contour and to obtain a smoothed contour. Then, in order to eliminate over-estimation and under-estimation errors, an additional smoothing algorithm is employed in which fundamental frequency values close to a constant multiplication of the dominant fundamental frequency value are corrected according to the dominant fundamental frequency as they are assumed to be estimation errors and are therefore corrected according to the dominant fundamental frequency value.

Considering the algorithm accuracy, the resulting time resolution is $\delta = \frac{1}{Fs} = 0.0625sec$ for $Fs = $16kHz. If the true fundamental frequency is 250Hz, the maximum quantization error would be approximately $\frac{T}{2P^2} = 1.95Hz$, where $T = \frac{1}{Fs}$. 

It should be noted that for the common application of cry classification, the exact value of the fundamental frequency is not of great interest, but rather the fundamental frequency contour (``pitch contour''). Thus interpolation, which is usually performed in the standard SIFT, is unnecessary here.  

\section{Results}
\label{sec:RESULTS}
Fig. \ref{fig:fig1} presents some typical results obtained using the modified SIFT. The cry segments in the examples shown in this figure were taken from one healthy infant.     
Note that for the unvoiced segment, the values of the autocorrelation sequence are much lower than the threshold 0.4, and indeed no fundamental frequency is estimated and the segment is classified as unvoiced. In cases (b) and (c), a regular phonated segment is presented in which the fundamental frequency is estimated as 380Hz and 383Hz, respectively. In case (d) which shows a hyperphonated cry segment, the fundamental frequency is much higher than in the previous cases.

\fig{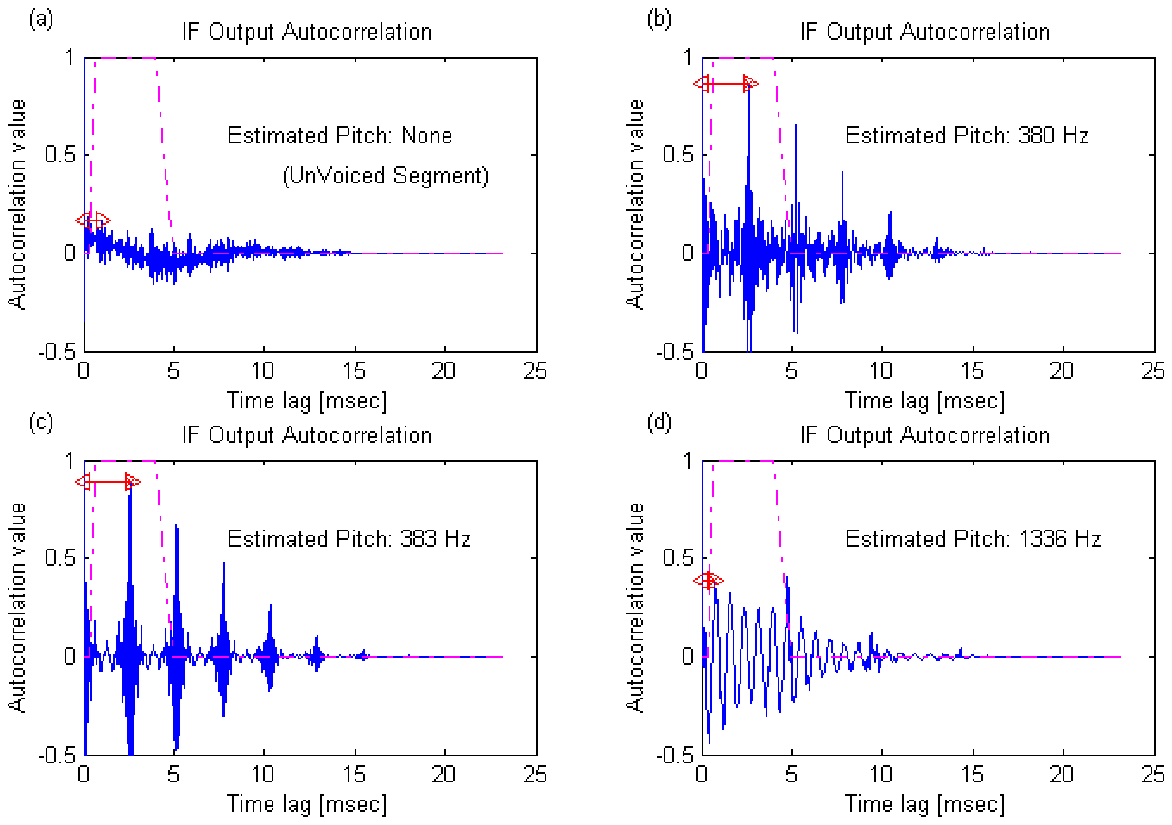}{8.0cm}{8.0cm}{Examples for the inverse filtering output autocorrelation signal and the estimated fundamental frequency.
The following cases are presented: (a) Unvoiced segment;  (b) Phonated voiced segment;  (c) Phonated voiced segment;  (d) Hyperphonated segment.}
      {fig:fig1}

Fig. \ref{fig:fig5} shows typical segments of cries for three medical states: healthy (left), cri-du-chat (middle) and hydrocephalus (right). The first line, marked as (a), shows time domain plots of the signals. The second line (b) shows spectrograms of the signals from which the fundamental frequency can be visually estimated. The spectrograms were estimated using 20msec frames of the signals, with 10msec overlap and an Hamming window. The third line (c) presents the fundamental frequency contours obtained using the modified SIFT. 
The figure demonstrates that there are major differences in the time and frequency domains among the three medical states.  The results indicate that the fundamental frequency in the two pathological states (cri-du-chat syndrome and hydrocephalus) is higher on the average than that of the healthy infant. This is in agreement with previous studies, in which the average fundamental frequency of infants with cri-du-chat was found to be 860Hz \cite{Vuorenkoski1966}\cite{Michelsson1980}, and the average fundamental frequency of infants with hydrocephalus was approximately 620Hz, with minimum and maximum values of 430Hz and 750Hz, respectively\cite{Michelsson1980}. Some fundamental frequency over-estimation errors occurred especially in the two pathological cases where rapid variations of the fundamental frequency contour exist. These errors were easily corrected by the smoothing algorithm, yielding flat or almost flat fundamental frequency contours, in accordance with the spectrogram. 

\fig{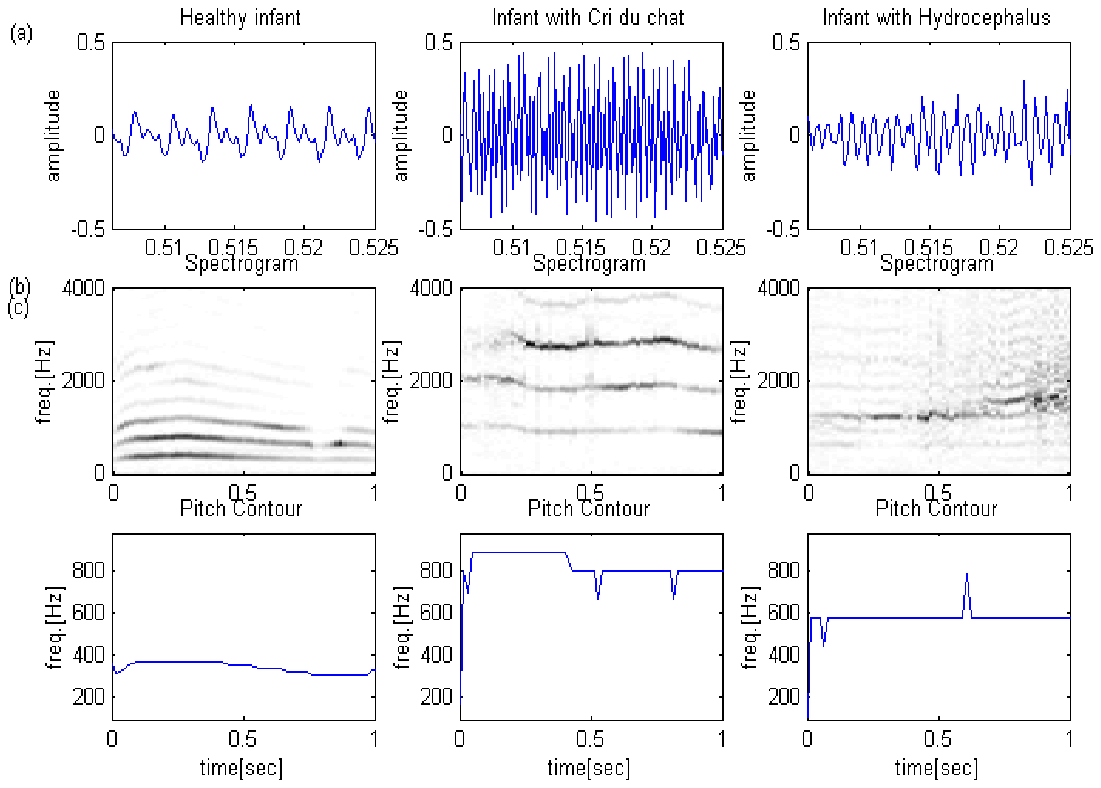}{8.0cm}{8.0cm}{Typical segments of cries for three medical states. Left: healthy; Middle: cri-du-chat; Right: hydrocephalus. 
(a) Time domain plots (b) Spectrograms (c) Fundamental frequency contour.}
      {fig:fig5}

Fig. \ref{fig:fig2} shows some typical results for the case of infants born with cleft palate. The left column represents the time-domain waveform, the spectrogram and the fundamental frequency contour, for infants without a palatal plate, while the right column represents the corresponding signals for the same infant with a palatal plate (see details in \cite{Lederman2008}).

\fig{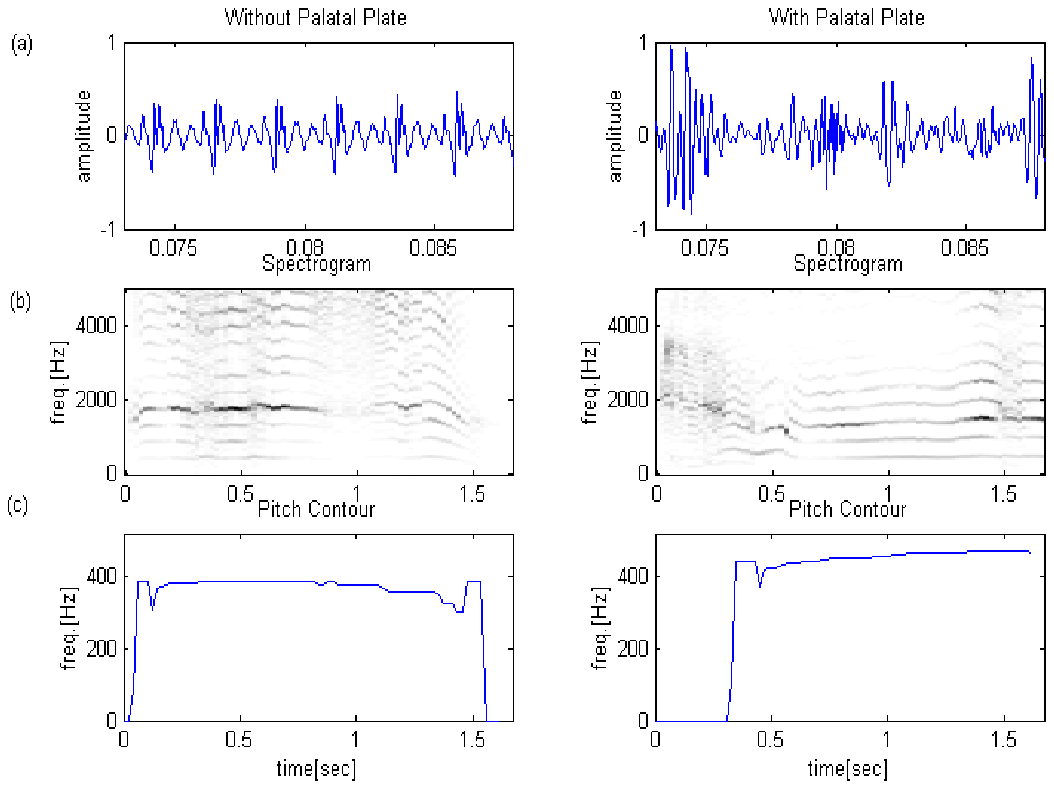}{8.0cm}{8.0cm}{Typical segments of cries of an infant with cleft palate. Left: infant without a palatal plate; Right: infant with a palatal plate.  (a) Time domain plots   (b) Spectrograms   (c) Fundamental frequency contour.}
      {fig:fig2}
 
Differences between the two medical states in the time domain graphs can be easily noticed, whereas in the spectrograms, as well as in the fundamental frequency contours, no significant difference exists. The average fundamental frequency in both cases is around 450-500Hz, only slightly more than for healthy infant (shown in Fig. \ref{fig:fig5}). This is also in accordance with previous reports on infants with cleft palate, which did not show any major differences between the fundamental frequency of infants with cleft palate and the fundamental frequency of healthy infants.

It should be noted that the smoothing algorithm plays an important role here since in some of the segments the SIFT algorithm tends to incorrectly estimate the fundamental frequency, in particular when abrupt changes of the fundamental frequency occur. 
Fig. \ref{fig:fig1} and \ref{fig:fig2} also show that the cries are almost entirely voiced (estimated fundamental frequency greater than zero). This observation is in accordance with previous results reported by Cohen and Zmora \cite{Cohen1984}. 

In order to evaluate the fundamental frequency estimation accuracy with the modified SIFT algorithm, the estimated fundamental frequency was compared to the ``real'' fundamental frequency estimated by visual inspection of the signals in the time domain. The error between the two values was calculated as a percentage of the visually-estimated fundamental frequency, as follows:
 			\begin{equation}
	 			\epsilon = \frac{1}{J} \sum_j 100 \frac{|\tilde{f_j} - f_j|}{f_j}
 			\end{equation}
where $\tilde{f}_i$  represents the estimated fundamental frequency and $f_j$  represents the ``real''  fundamental frequency, for the $j^{th}$ frame. The error was calculated separately for hyperphonated segments ($f_j \geq$ 700Hz) and phonated segments (250Hz $\leq f_j \leq $700Hz). Table \ref{table:tab1} presents the averaged errors, calculated over 100 frames for each segment type.

\begin{table}[ht]
\caption{Fundamental frequency estimation error} % title of Table
\centering % used for centering table
	 
\begin{tabular}{|c|c|} % centered columns (4 columns)
\hline %inserts double horizontal lines
    Segment type	&  Error rate(\%)\\
\hline
    Hyperphonated  & 6.15 \\
\hline
    Phonated 			 & 3.75\\
\hline 
\end{tabular}
\label{table:tab1} % is used to refer this table in the text
\end{table}

Both errors are relatively low. Furthermore, as expected, the average error is higher in the case of hyperphonated segments due to widening of the harmonic structure in those segments. 

\section{Conclusions}
\label{sec:SUMMARY}
In this paper, a modified SIFT algorithm for cry fundamental frequency estimation is presented and investigated using real data. It is shown via some typical examples of pathological and healthy cry signals that the modified SIFT may be used to reliably estimate the fundamental frequency, overcoming the problem of under-estimation and over-estimation. The performance of the algorithm is evaluated  comparing to visually-estimated fundamental frequency, and is shown to yield an estimation error of around $3.75\%$ and $6.10\%$, for phonated and hyperphonated cry segments, respecively. The modified SIFT algorithm is an efficient and simple method to estimate the fundamental frequency of infants' cry and to perform voiced/unvoiced segmentation. 

%Maximum a-posteriori probability pitch tracking in noisy environments using harmonic model
%Tabrikian, J.; Dubnov, S.; Dickalov, Y.
%Speech and Audio Processing, IEEE Transactions on
%Volume 12, Issue 1, Jan. 2004 Page(s): 76 - 87
%Digital Object Identifier   10.1109/TSA.2003.819950
%Summary: Modern speech processing applications require operation on signal of interest that is contaminated by high level of noise. This situation calls for a greater robustness in estimation of the speech parameters, a task which is hard to achieve using standard speech models. In this paper, we present an optimal estimation procedure for sound signals (such as speech) that are modeled by harmonic sources. The harmonic model achieves more robust and accurate estimation of voiced speech parameters. Using maximum a posteriori probability framework, successful tracking of pitch parameters is possible in ultra low signal to noise conditions (as low as -15 dB). The performance of the method is evaluated using the Keele pitch detection database with realistic background noise. The results show best performance in comparison to other state-of-the-art pitch detectors. Application of the proposed algorithm in a simple speaker identification system shows significant improvement in the performance.

%\nocite{Tabrikian2004}

\section*{Acknowledgment}
% optional entry into table of contents (if used)
%\addcontentsline{toc}{section}{Acknowledgment}
    This research is dedicated to the memory of Prof. Arnon Cohen who initiated the cry analysis and classification project.

\bibliography{dror}
\bibliographystyle{IEEEtran}

\end{document}

%% file: myshorts.tex
\def\0{{\mathbf 0}}

\def\beta{{\mathbf \eta}}

\newcommand{\fig}[5]
{
\begin{figure}[htb]
  \centering
  \includegraphics[width=#2,height=#3]{#1.eps}
  \caption{#4}
  \label{#5}
\end{figure}
}